\documentstyle[11pt]{article}

\advance \voffset by -0.8in
\advance \hoffset by -0.4in
\newcommand{\ix}[1]{{\mbox{\tiny #1}}}
\newcommand{\bd}[1]{\mbox{\boldmath$#1$\unboldmath}}
\def\be{\medskip \begin{equation}}
\def\ee{\medskip \end{equation} }
\def\bmphi{\mbox{$$\boldmath$ \phi $\unboldmath$$}}
\def\bphi{\mbox{\boldmath$\phi$\unboldmath}}
\def\cp{{\cal CP}}

\begin{document}


\begin{flushright}
\parskip 4pt
DTP 95-75

April 1996
\end{flushright}

\begin{center}

{\huge Topological Chern--Simons vortices in the $O(3)\,\sigma$--model\\}
\vspace{1cm}
{\Large J. Gladikowski\footnote{e-mail: {\tt Jens.Gladikowski@durham.ac.uk}}
\medskip\\}
{\it Department of Mathematical Sciences\\ South Road, Durham DH1 3LE, England}

\vspace{1cm}

{\large {\bf Abstract}}
\end{center}

We present a classical, gauged $O(3) \, \sigma$--model with
an abelian Chern--Simons term.
It shows topologically stable, anyonic vortices as solutions.
The fields are
studied in the case of rotational symmetry and analytic
approximations are found for their asymptotic behaviour.
The static Euler--Lagrange equations are solved numerically, where 
particular attention is paid to the
dependence of the vortex' properties
on the coupling to the gauge field.
We compute the
vortex mass and charge as a function of this coupling and
obtain bound states for two--vortices as well as 
two--vortices with masses above the stability threshold.


\section{Introduction}

The $O(3)\, \sigma$--model in 2--dimensional Euclidean space
is a classical  field theory which supports soliton solutions
\cite{bep}.
Its scale invariance can be broken by the addition of a potential term.
This does not prevent the soliton from shrinking, however, its size
can be fixed by the inclusion of higher order terms in the field gradient
\cite{sky}.
An example for such a theory is the baby Skyrme model
\cite{multi}.
Alternatively, the scale invariance of the $O(3)\, \sigma$--model
can be removed and the soliton
be stabilised (at least in principle)
by gauging a $U(1)$ subgroup of the fields internal symmetry group
\cite{nar,Bernd,jps}.
The dynamics of the $U(1)$ gauge field in such models is ruled by
Maxwell--  and/or Chern--Simons actions. 
For each of these cases 
potential terms have been constructed such that the 
corresponding models  yield  self--dual equations of Bogomol'nyi type.
The potential term determines the asymptotic behaviour of the fields
which can either obey the gauge symmetry \cite{Bernd,ghosh}
or break it \cite{lee1}.
The models with broken gauge symmetry show topological soliton solutions 
with quantised magnetic flux and in this they resemble the well--studied
vortices in the abelian Higgs model and its generalisations
, see for instance \cite{jlw,llmk}.

Here we investigate a gauged $O(3) \,\sigma$--model with Chern--Simons 
action. 
Chern--Simons theories are an object of intense research 
because their quantised version is relevant
for  systems of strongly correlated 
electrons e.g. in superconductors or in the quantum Hall effect
\cite{wil}.
In this paper we consider a static classical Chern--Simons model,
whose potential term preserves the gauge symmetry and
is chosen to  produce 
exponentially localised configurations. 
They carry fractional 
angular momentum and have a lower topological bound on the energy which is,
however, not saturated. 
We solve the equations of motion
numerically for radially symmetric fields
and study the dependence of the solutions
on the coupling strength
to the gauge field. 
We also look at two vortices on top 
of each other and on their mutual attraction dependent on their coupling.
The asymptotic behaviour of the fields is studied analytically
and conclusions about intervortex forces are drawn.

Recently, static solitons were found 
in a gauged $\cp^1$ model which includes
a Chern--Simons term  and a potential term equivalent to
the one considered here \cite{btw}.
In its standard version the $\cp^1$ model represents 
merely a different choice of fields to the $O(3) \,\sigma$--model.
In \cite{btw}, however, 
the gauged symmetry is the internal $U(1)$ symmetry of the
two--component complex $\cp^1$ vector which lies on ${S}^3$.
Therefore we expect our solutions to be different 
to the ones presented in \cite{btw}, but it is nevertheless instructive
to compare them.


\section{Chern-Simons solitons revisited}


We consider the following  Lagrangian of a gauged  $O(3)\,\sigma$-model
in (2+1) dimensions.
It contains a potential term and the behaviour of the 
gauge field $A_\alpha$ is governed by a Chern--Simons term
\be
\label{lagr}
{\cal L} = \frac{1}{2}\left(D_\alpha \bmphi\right)^2 
    -\frac{\kappa}{2}\epsilon^{\alpha \beta \gamma}\partial_\alpha A_\beta A_\gamma -
            \mu^2(1-{\bf n}\cdot \bd{\phi}) \,.
\ee
The fields $\bmphi$ are three-component real vectors and subject to the
constraint $\bmphi \cdot \bmphi=1$, hence they take values on 
the two-sphere $S^{2}_{\phi}$.
The metric is chosen to be $ g_{\alpha \beta} = $ diag$(+, -, -)$.
Throughout this paper, greek
indices run from 0 to 2 while Latin indices denote
the two spatial dimensions 1,2.
We work in geometrical units in which the velocity of light $c=1$. 
$\kappa$ and $\mu$ are real coefficients of dimension length and 1/length
respectively and for dimensional reasons the Lagrange density (\ref{lagr})
should be thought of being multiplied by an overall factor of
dimension energy.
We fix our mass scale by putting this factor to one.
The fields $\bmphi$ will be frequently referred to as matter--fields
(in distinction to
the gauge fields) and to  $S^{2}_{\phi}$ as the iso--space.
The potential term in (\ref{lagr}) reduces the symmetry 
of the model to $O(2)_\ix{iso}$, i.e.
to rotations and reflections perpendicular to the vector {\bf n}.
It is this symmetry that is to be gauged and 
by choosing ${\bf n} = (0,0,1)$ we select the $SO(2)_\ix{iso}$ subgroup 
which consists of unimodular rotations about the $z$--axis.
$D_\alpha (\bmphi)$ is the covariant derivative and given by:
\be
D_\alpha \bmphi = \partial_\alpha \bmphi + A_\alpha ({\bf n} \times \bmphi)\,.
\ee
The ungauged Lagrangian  shows symmetry under combined reflections in
space and iso--space:
\be
P:\quad(x_1, x_2) \rightarrow (-x_1, x_2) \qquad\mbox{$$ and $$}
\qquad C:\quad(\phi_1 , \phi_2 ) \rightarrow
(-\phi_1, \phi_2)\,,
\ee
which can be thought of as a parity operation and charge conjugation. The
Chern--Simons term breaks the parity symmetry explicitly 
by changing its sign under $P$.
It also breaks the
time--reflection symmetry $T$ which corresponds to $A_0 \to -A_0$.
However, the Lagrangian is still symmetric under $CPT$.

The potential term can be thought of physically as an analogue 
to the Zeeman coupling between   
spin fields $\bmphi$ and an external, 
constant magnetic field in {\bf n}--direction with coupling strength
$\mu^2$.
Such terms occur for example naturally in the description of the
quantum Hall effect.

Because we are interested in configurations
with finite energy, we require
that the potential term and the  covariant derivative
vanish at spatial infinity.
Hence we impose:
\label{phibound}
\be
\lim_{r \to \infty} \bmphi (r) = {\bf n}\,.
\ee
This boundary condition allows to one--point compactify the physical space
such that fields $\bmphi$ are maps:
\be
\bmphi \quad : \quad S^{2}_\ix{x} \rightarrow  S^{2}_\phi\,.
\ee
These maps are 
elements of homotopy classes which form a group isomorphic to
the group of integers.
This integer or degree $N$ counts the number of times $ S^{2}_\phi$ is 
covered by a single covering of $ S^{2}_\ix{x}$. It can be written
as the integral over the zero component of the
topologically conserved current:
\be
\label{o3curr}
l_\alpha = \frac{1}{8\pi}\epsilon_{\alpha \beta\gamma}
\bmphi \cdot(\partial_\beta \bmphi  \times  \partial_\gamma \bmphi)\,,
\ee
such that the degree  $N$ is obtained from
\be
\label{degree}
N =\int d^2 x \, l_0\,,
\ee
where the range of integration is $S^{2}_\ix{x}$.
Following the standard convention we will call
finite energy solutions with
$N >0$ vortices and those with $N<0$ antivortices.
(\ref{degree}) is in fact the topological charge of the $O(3)\,\sigma$ fields and
it is not obvious that this is also a topologically conserved quantity
in the gauged model.
We will therefore address this question again below.

The equations of motion derived from (\ref{lagr}) 
can be written in terms of the matter-current $J_\alpha$
and the electromagnetic current $j_\alpha$
\be
\label{J}
{\bf J}_{\alpha} =
\bmphi \times D_{\alpha} \bmphi \quad , \quad 
j_{\alpha} ={\bf n}\cdot {\bf J}_{\alpha}\,.
\ee
The Euler-Lagrange equations are
\begin{eqnarray}
\label{eom}
D_{\alpha}{\bf J}^{\alpha} = &\mu^2({\bf n}\times \bphi)\\
\label{eom_j}
j_{\alpha} = &-\kappa\epsilon_{\alpha \beta \gamma}\partial^\beta A^\gamma\,.
\end{eqnarray}
Note that by (\ref{eom_j})
the gauge fields are completely determined by first order 
equations and do not have own dynamics in the strict sense.
Equation  (\ref{eom_j}) for $\alpha = 0$ is Gauss' law
\be
\label{Anought}
D_0\bmphi = 
-\frac{\kappa B ({\bf n} \times \bmphi)}{ 1-({\bf n}\cdot \bmphi)^2 } \,,
\ee
where we have used that ${\bf n} = (0,0,1)$ and $B =\epsilon_{0 i j}\partial^i A^j$,
taking $\epsilon_{0 1 2} =1$.
The equation of motion  (\ref{eom_j}) implies that
for non--singular $A_\alpha$ the
electromagnetic current $j_\alpha$ 
is conserved   ($\partial_\alpha j^{\alpha} =0$).
The current can be written conveniently  
as $j_\alpha = (\rho, j_i)$, where $\rho$ is the charge density of the
soliton while $j_i$ denotes its electric current.
The Lagrangian (\ref{lagr}) can be expressed in terms of $j_\alpha$:
\be
{\cal L} = \frac{1}{2}\left(D_\alpha \bmphi\right)^2 
           -\frac{1}{2}A_\alpha j^\alpha +
            \mu^2(1-{\bf n}\cdot \bd{\phi}) \,.
\ee
This shows explicitly  that the gauge fields $A_\alpha$ are coupled 
to the electromagnetic current $j^\alpha$.
The electric field ${\bf E}$ 
and the magnetic field $B$ are related to $j_\alpha$ as follows:
\be
\label{elmag_fields}
B = -\frac{\rho}{\kappa}\quad , \quad E_i = \epsilon_{i j} \frac{j^j}{\kappa}\,.
\ee
The first equation leads to 
a relation between the magnetic flux $\Phi$ and the electric charge $Q$ of the
configuration 
\be
\label{fl_charge}
\Phi =
\int \, d^2 x \,B = 
-\frac{1}{\kappa}\int \, d^2 x \,\rho = -\frac{Q}{\kappa} \,.
\ee
The theories energy--momentum tensor 
is obtained by the
variation of the Lagrangian with respect to the metric $g_{\alpha \beta}$
\be
T_{\alpha \beta} = (D_\alpha  \bphi)(D_\beta  \bphi) -
g_{\alpha \beta}\left( \frac{1}{2}(D_\gamma \bphi)(D^\gamma \bphi) -
\mu^2(1-{\bf n}\cdot\bphi)\right)\,.
\ee
The integral over the component $T_{00}$ is the
total energy of the soliton
\be
\label{en}
E_\ix{CS}[\bphi, A] =\\
           \int d^2 x \, 
           \frac{1}{2}\left(D_0 \bmphi\right)^2 +\frac{1}{2}\left(D_i \bmphi\right)^2+
          \mu^2(1-{\bf n}\cdot \bd{\phi}) \,.
\ee
Note that the Chern--Simons term does not contribute directly
to the energy because of 
its metric independence.
The rotational symmetry of the Lagrangian
leads to a conserved angular momentum ${\bf M}$ of the soliton
\be
\label{ang_mom}
{\bf M} = \int \, d^2 x \,\, ({\bf x}\times {\bf p}),
\ee
where the cross product stands for $x_1p_2-p_1x_2$.
${\bf M}$  is a vector pointing perpendicular out of the plane of motion.
The components of the momentum density {\bf p} are given by
\be
p_i = T_{ 0 i } = D_0 \bmphi \cdot D_i \bmphi\,.
\ee
%


\section{A bound on the energy}


Next we give a proof that $E_\ix{CS}[\bphi, A]$,
the  energy in our model given by (\ref{en}), is 
bounded from below by a topologically conserved quantity.
This is not obvious, because
the gauged pure $O(3)\,\sigma$--model does not have
a lower bound on the energy, unlike its ungauged counterpart,
where the solutions saturate the Bogomol'nyi limit.
The first step in the proof is to use an auxiliary energy
functional $E_\ix{aux}[\bphi, A]$ which 
is of Bogomoln'nyi type
and was  constructed in \cite{Bernd}.
Because we wish this section to be self--contained, we
will repeat below parts of the  analysis given in this reference.
First, we show that the energy gap between  $E_\ix{CS}$ and $E_\ix{aux}$
(or a multiple of it) is positive and then
complete the argument by demonstrating
that $E_\ix{aux} \geq 4\pi|N|$.
$E_\ix{aux}$ reads as:
\be
\label{aux0}
E_\ix{aux}[\bphi, A] = \frac{1}{2}\int d^2 x \, 
           B^2+\left(D_i \bmphi\right)^2+
           (1-{\bf n}\cdot \bmphi)^2 \,.
\ee
In order to be consistent in the notation of dimensions,
both the potential term and the magnetic field
must be thought of being multiplied 
by a parameter of dimension 1/length squared and length 
respectively.
These parameters are of magnitude one and subpressed in (\ref{aux0}).
To compare $E_\ix{CS}$  with $E_\ix{aux}$ one first observes that
$(D_0 \bmphi)^2 \geq \kappa^2 B^2$, due to Gauss' law
(\ref{Anought}).
Now we carry out a rescaling of $x$ in our functional $E_\ix{CS}$,
namely $x \to \kappa x$, which transforms $B \to B/ \kappa^2$ and
$\bmphi(x) \to \bmphi(\kappa x)$.
The potential term then reads as 
\mbox{$\kappa^2 \mu^2(1-{\bf n}\cdot \bmphi)$} and is 
greater than $(1-{\bf n}\cdot \bmphi)$ if $\kappa  \geq 1/\mu$.
To verify that $E_\ix{aux}$ is smaller than $E_\ix{CS}$ we use that
since $0 \leq (1-{\bf n}\cdot \bmphi) \leq 2$, it
follows that 
$(1-{\bf n}\cdot \bmphi) \geq \frac{1}{2}(1-{\bf n}\cdot \bmphi)^2$ and
one sees that  for $E_\ix{aux}$ holds
\be
\label{aux1}
E_\ix{CS}\geq E_\ix{aux} \quad \mbox{$$if$$}\quad \kappa \geq 1/\mu\,.
\ee
In the case  $\kappa <$1/$\mu$  we 
assess an energy bound by multiplication of each individual term 
in the energy density with $\kappa^2 \mu^2$. This gives 
\be
E_\ix{CS} \geq
\kappa^2\mu^2 E_\ix{aux} \quad \mbox{$$if$$}\quad \kappa < 1/\mu\,.\\
\ee
This already proves the bound for $E_\ix{CS}$, but 
it is instructive to see in detail that  $E_\ix{aux}$ defines a Bogomol'nyi
model.
In order to achieve this, we rewrite the auxiliary energy
functional as
\be
\label{aux2}
E_\ix{aux}[B, \bd{\phi}] =
\frac{1}{2}\int d^2 x\, \left( D_1 \bmphi \pm  
\bmphi \times  D_2 \bmphi \right)^2 + 
\left( B \mp (1-{\bf n}\cdot \bd{\phi})\right)^2 \pm
\int d^2 x\, L_0\,.
\ee
$L_0$ is composite of the cross terms  and can be understood as the zero
component of the solitons gauge invariant topologically conserved 
current:
\be
L_\alpha = \epsilon_{\alpha \beta \gamma} 
\left(
\bmphi \cdot (D^\beta \bmphi \times D^\gamma \bmphi) +
\partial^\beta A^\gamma  (1-{\bf n}\cdot \bd{\phi}) 
\right)\,.
\ee
Up to a surface term, this current is 
equivalent to $l_\alpha$, 
the topological current of the
ungauged $O(3) \,\sigma$--model  (\ref{o3curr}).
If the solutions are required to have finite energy, then
$\bd{\phi}$ must tend to zero 
faster than $1/r$ as $r$ goes to infinity, hence it follows by 
Stokes' theorem that the surface term  integrates to zero.
It was pointed out in \cite{lee1}
that the conserved topological charge (integral over $L_0$) equals
the degree $N$ of the map $\bmphi$ if the gauge symmetry is unbroken
(as it is in our case) but can differ from $N$ for broken symmetry.
In order to saturate the Bogomol'nyi bound,
both squares  in (\ref{aux2}) have to vanish, such 
that the following two (anti--) self--dual equations are  read off
\be
D_1 \bmphi =\mp \bmphi \times  D_2 \bmphi \,\qquad,\qquad
B = \pm (1-{\bf n}\cdot \bd{\phi})\,.
\ee
These equations 
were discussed in detail for a special choice
of the fields in \cite{Bernd}.
There it was shown that they yield
a one--parameter family of solutions which are
degenerated in their energy but differ in their magnetic flux.

By using the sign ambiguity in front of the integral over $L_0$ in (\ref{aux2})
we can restrict our
discussion to the case  $B>0$ and the upper sign
without a loss of generality. 
Equation (\ref{aux2}) then implies
\be
E_\ix{aux}\geq \int d^2 x L_0 = 4\pi|N|\,,
\ee
The equality holds for self--dual solutions.
%

\section{Static vortex solutions}


To find static solutions in our model we restrict ourselfs to the 
two--dimensional hedgehog \cite{hedge},
which is in terms of the polar coordinates
$(r, \theta)$:
\be
\label{hedgehog}
\bmphi ( r, \theta ) =
\left(
\begin{array}{rl}
\sin f(r) & \cos  n \theta   \\
\sin f(r) & \sin n \theta  \\
\cos f(r) &
\end{array}
\right) \,. 
\ee
For this field the topological charge density, the integrand of
(\ref{degree}), equals 
\be
l_0 = \frac{n}{4\pi r}f^{'}\sin f\,,
\ee
where the prime denotes the derivative with respect to $r$.
By integration  one easily sees that $n=-N$.
For the gauge field $A_\alpha$, the most general ansatz which leads
to radially symmetric and static fields is given by
\be
A_0 = nv(r)\,,\quad A_\theta = na(r)\,,\quad A_r = h(r)t\,,
\ee
where $t$ denotes the time and the factor $n$ is introduced for
convenience.
We fix our gauge by putting $A_r=0$ and obtain the equations
\begin{eqnarray}
f^{''} + \frac{f^{'}}{r} &=& n^2\left(\frac{(a+1)^2}{r^2} -v^2 \right) \sin f \cos f +\mu^2\sin f \label{radeom}\\
\label{eom_v}
v^{'} &=&-\frac{1}{\kappa}\frac{(a+1)}{r}\sin^2 f\,.
\end{eqnarray}
Gauss' law (\ref{Anought}) reads in terms of $a$, $v$ and $f$:
\be
\label{eom_a}
a^{'} =-\frac{1}{\kappa} r v \sin^2 f \,.
\ee
We are interested in finite energy  configurations, which
requires that  $D_\alpha(\bd{\phi}) \to 0$ as $r \to \infty$.
To guarantee this and the regularity of the fields at the origin we 
impose the following boundary conditions
\be
\label{bound}
\begin{array}{rl}
&a(0) = 0\,,\quad f(0) = \pi\,,\quad v(0) = v_0\\
& \lim_{r \to \infty} f(r) = 0\,,\quad \lim_{r \to \infty} a(r) = a_\infty\,,
\quad \lim_{r \to \infty} v(r) = v_\infty
\end{array}
\ee
where $v_0,v_\infty$ and $a_\infty$ are constants.
With these boundary conditions it is clear that constant fields
$a$ and $v$ are not a solution of (\ref{eom_v}) and  (\ref{eom_a}),
which can be shown by contradiction.
If
$a$ were a constant it would have to be zero everywhere 
because of (\ref{bound}), in which case
(\ref{eom_v}) implies that 
$v$ is not a constant which in turn, 
via  (\ref{eom_a}) leads to a non--constant $a$.
A similar argument applies for the case of $v$ being constant.
Hence the Euler--Lagrange equations do not lead
to vanishing flux and charge.

The total (static) energy is given as the integral over the energy density $e$,
which reads in terms of the fields $f, a$ and $v$ (\ref{en})
\be
\label{led}
e = \frac{{f^{'}}^2}{2}+\frac{n^2}{2}\left(\frac{(a+1)^2}{r^2}+v^2\right)\sin^2 f
+\mu^2(1-\cos f)\,.
\ee
For the angular momentum  (\ref{ang_mom}) one obtains
\be
{\bf M} = -\pi\kappa N a_\infty(a_\infty + 2N){\bf n}\,.
\ee
Hence one sees that the angular momentum of the vortex is
fractional and the vortices are (classical) anyons.
The electromagnetic fields (\ref{elmag_fields}) are 
radially symmetric by construction and read as
\be
B = N \frac{a^{'}}{r}\quad , \quad E_r = Nv^{'}\,.
\ee
The electric charge and magnetic flux are not topologically quantised
(unlike in the abelian Higgs model, for instance)
and depend on the parameters in the model 
\be
\label{flux_charge_r}
\Phi = N\int \,rdr d\theta \,\frac{a^{'}}{r} = 
-2\pi N a_\infty = -\frac{Q}{\kappa}\,.
\ee 


\section{Asymptotics}


The boundary conditions (\ref{bound}) allow us to derive asymptotic
approximations to the equations of motion (\ref{radeom}).
By approximating $\sin f \approx f$ and $\cos f \approx 1$
for large $r$, the equation for $f$ simplifies  to
\be
\label{asymp_f}
f^{''} + \frac{f^{'}}{r} = 
n^2 \left(\frac{(a_\infty+1)^2}{r^2} +  k^2  \right)f \,,
\ee
where 
\be
\label{wavenumb}
k^2 =  \mu^2 - n^2v^2_\infty\,.
\ee
The asymptotic solution of $f$ depends on the value  $k$ takes.
There are three possible cases.

1.) $|\mu| > |nv_\infty|$, $k$ real.\\
The solution to (\ref{asymp_f})  for real $k$
are given by  modified Bessel functions $f \sim K_m(kr), m
=n(a_\infty+1)$  with the asymptotic behaviour
\be
\label{f_large}
f \sim \frac{1}{\sqrt{r}} e^{-kr}\,, \quad 
\ee
This shows that $k$ can be understood as the
effective mass of the matter fields $\bmphi$,
by denoting the inverse decay 
length of the profile function. 
The asymptotics of the field are determined by the  potential term
which defines the theories vacuum structure. 
Therefore it is not a surprise that
the vortex' matter field looks  asymptotically like  the
baby Skyrmion investigated in \cite{multi}, where the same potential 
term was used.

2.) $\mu < |nv_\infty|$, $k$ imaginary.\\
This case leads to oscillating fields with an amplitude that
falls off proportionally to $1/\sqrt{r}$ in leading order.
The substitution $\tilde{k} = ik$ in (\ref{f_large})
verifies this instantly and also shows that $k$ is proportional
to the inverse wavelength of the oscillations. 
The energy density of these solutions behaves  asymptotically 
like $1/r$ in leading order and hence the 
energy of these fields is infinite. 
This is,  of course, not a physically relevant solution
such that we  exclude it from our further discussion.

3.) $\mu = |nv_\infty|$, $k=0$.\\
The critical case is in fact just a
special case of 1.), with vanishing exponential such that
the profile function $f\sim 1/r$. The energy of these solutions is also
infinite, because the leading term in the energy density is proportional
to $v_\infty^2f^2$.
Numerically we find that all these solutions occur but restrict
our discussion to the case 1.), which gives the following constraint
on the solutions: 
\be
|\mu| > |nv_\infty|.
\ee
Using expression (\ref{f_large}),
we find for the electric and magnetic field
in the limit of large $r$
\be
E_r \sim \frac{1}{r} e^{-2k r}\,,\quad B \sim \frac{1}{r} e^{-2k r}\,.
\ee
This  shows that the electromagnetic fields fall off 
much faster than the matter field $f$. 
Therefore the electromagnetic interactions
are expected to be negligible in the
context of long--range vortex interactions. 
The electric field is a vector lying in the plane
of motion while the magnetic field $B$  can be thought
of as pointing  perpendicular out of the plane of motion.
Its asymptotic shape is similar to the one of the 
Skyrme--Maxwell soliton discussed in \cite{jps}, where it 
was argued that such a magnetic field resembles a
magnetic dipole in two--dimensional electrodynamics.

For small $r$ the fields can be approximated by power series
\be
\label{smallr}
f \approx \pi + cr^{|n|}  \,, \quad 
v \approx v_0 + dr^{2|n|}  \,, \quad
a \approx gr^{2|n|+2} \,,
\ee
where $c$ and $v_0$ are free parameters while $d$ and $g$ are given as functions of
$n, \kappa, c$ and $v_0$.
Note that for finite energy 
solutions $c$ and $v_0$ are not completely independent on each other.

For the Skyrme--Maxwell solitons it was found that the electromagnetic 
short range interaction decreases the energy per soliton
and in particular leads   to  more strongly bound two--soliton states.
Here, having a non--zero electrical charge distribution
we expect this effect to be weakened by the Coulomb repulsion of the solitons
electric field.


\section{Numerical results}


We solved the set of equations (\ref{radeom}) numerically by
using a shooting method and a relaxation method.
For both the shooting method and the time evolution in the relaxation method
we employed a fourth--order Runge--Kutta method.
In order to perform the numerical integration
we had to fix the parameters in our model.
Using geometric units in which the
energy and length are of unit one, we are left
with $\mu$ and $\kappa$ to be fixed.
However, the parameter space is in fact
one--dimensional which can be verified by 
carrying out the rescaling $x \to \kappa x$, $B \to B$/$\kappa^2 $.
Thus we can fix $\mu$ for all our computations
without a loss of generality.
We choose $\mu =\sqrt{0.1}$, a value which 
allows us to compare our numerical results with
the ones obtained 
in the Skyrme--Maxwell model \cite{jps} 
where the same value has been used.

We looked at the dependence
of the solutions of degree $N=1$ and $N=2$ for a range of $\kappa$. 
This parameter determines the strength of the Chern--Simons term
and is proportional to the square root of the inverse coupling to the
gauge field, which can be seen by a simple 
substitution $A_\alpha \to A_\alpha /\sqrt{\kappa}$. \\ 
Fig. 1a) shows the dependence of the static energy or mass on $\kappa$.
Small $\kappa$ which corresponds to strong coupling leads to
lighter vortices for both the one--vortex and the two--vortex.
For large $\kappa$ the mass or static energy $E_\ix{CS}$
tends to a constant but
remains relatively close to the Bogomol'nyi bound, staying below 1.1
(in units of $4\pi|N|$) for the one--vortex and the
two--vortex. Thus our vortices are significantly  lighter
than the gauged baby Skyrmions, which tend to a mass of
$E_\ix{SM}=1.546$ for weak coupling.
The energy gap arises partly 
due to the  Skyrme term which is not present here.

It is particular interesting to look 
at the relative static energy per vortex.
We denote the energy of the one (two) vortex by $E^1 \, (E^2)$ .
The energy  difference $\Delta E = E^2 - 2E^1 $ can be interpreted 
as binding and excess energy  of the two--vortex for $\Delta E < 0$
and $\Delta E > 0$ respectively.
In the case $\Delta E < 0$
the vortices form bound states while for
$\Delta E > 0$ we expect
that vortices on top of each other are unstable under perturbations and
experience a repulsive force. From Fig. 1a) 
it is clear that in our model both cases occur.
For small $\kappa$ the two--vortex is in an attractive regime
as it is for large  $\kappa$, however, there is an intermediate
region
$\kappa_\ix{cr}^\ix{l} < \kappa < \kappa_\ix{cr}^\ix{h}$
for which the 
two--vortex is unstable (in the sense that its decay is
energetically favourable).
$\kappa_\ix{cr}$ is the critical coupling for which $\Delta E =0$.
Numerically we find that $\kappa_\ix{cr}^\ix{l}=0.632$ and 
$\kappa_\ix{cr}^\ix{h}=2.215$.

This result can be explained in a semiqualitative way.
In the limit of large $\kappa$ the gauge fields decouple from
the matter fields and
become very small as compared to the matter fields.
The study of ungauged solitons (e.g. in \cite{multi}) showed 
that pure matter forces are often attractive
for two--solitons.
This is also the case in our model.
For very small $\kappa$, however, the magnetic flux tends to a constant
(the reader finds  the explanation for this below)
while the repulsive Coulomb force is 
$ \propto Q^2 \propto \Phi^2\kappa^2$, thus becoming weaker with 
increased coupling, see Fig. 1b). 
We found numerically that increased coupling leads to a 
stronger bound configuration and in this our model is similar to
the Skyrme--Maxwell model.
The intermediate or repulsive range can be understood as 
a regime in which the Coulomb repulsion dominates the attractive forces
of matter and magnetic field. 
It is within this range that the electric charge has its maximum value
$Q_\ix{max} = Q(\kappa_\ix{max})$, where numerically $\kappa_\ix{max}= 0.75$
(for $N=1$) and  $\kappa_\ix{max}= 0.92$ ($N$=2).

In showing such a behaviour, the vortices resemble the
fields of the abelian Higgs model where a similar transition
between repulsive and attractive regime occurs, depending on the 
strength of the potential term.
This characteristic is used to describe temperature driven
phase transitions between type-I and type-II superconductors.
The shape of the vortex shows a strong dependence on $\kappa$,
which is foreseeable by the interpretation of $\kappa$ as the
coupling parameter to the gauge field.
Because the vortex' Coulomb interaction is repulsive it
favours a spreading of the soliton.
In agreement with this picture, we obtain vortices with 
their maximum width  range at $\kappa\approx 1$, where the 
electric charge takes its maximum values.
See also Fig. 2a) and
2b), where the energy density $e$ and the
profile function $f$ are plotted in dependence on the coupling.

On the other hand, 
if the electromagnetic interaction is coupled only very weakly , the
Lagrangian reduces to the $O(3)\,\sigma$--model term plus the potential term, 
such defining a configuration which is known to be unstable
against shrinkage 
due to the Hobart--Derrick theorem \cite{der}.
In accordance with this discussion one sees from 
Fig. 2 that for large $\kappa$ 
the vortex becomes more localised.
This clearly shows that the potential term in the Lagrangian favours 
a shrinkage of the vortex.

We also observed that 
increased magnetic flux as it occurs for small $\kappa$
leads to more localised solitons, like it does 
for Skyrme--Maxwell solitons \cite{jps}.
For very small $\kappa$ both 
the gauge fields $a$ and $v$ tend to singular 
configurations at the origin. 
In this limit, the gauge field $v$ takes large values at the origin and is
zero everywhere  else, while
$a$ tends to $-1$ everywhere except at $r=0$, which is fixed by 
its boundary condition, see Fig. 2c) and 2d).
In that the behaviour of $a$ is similar to the gauge field in the
Skyrme--Maxwell model. 
We conjecture that the origin of this coincidence is the
particular ansatz chosen for the gauge field, which leads
to terms in the energy density  depending on $(a+1)^2$ 
and so makes the
value $a_\infty =-1$ exceptional.
The strong coupling limit therefore leads to  dynamically quantised
flux  and in addition implies via
(\ref{flux_charge_r}) that the 
electric charge vanishes for $\kappa \to 0$.
For the electric and the magnetic field we find 
that they form a ring (cf. Fig. 3b), a feature 
which was also observed by 
Jackiw and Weinberg in a self--dual Chern--Simons model \cite{jlw}
(where the matter fields are complex scalar).
For Skyrme--Maxwell solitons, toroidal configurations
were seen only for topological charge two.

We also looked at the vortex' shape for $N=2$
and its dependence on $\kappa$.
The two vortex has the shape of a ring for all $\kappa$,
a  picture not unfamiliar in  planar soliton theories. 
For equivalent coupling, the fields of the two--vortex
decay slower than those of the one--vortex. 
This can be understood by looking at formula (\ref{wavenumb}).
Our numerical results show  that
$v_\infty$ depends strongly  on the coupling $\kappa$ but only weakly
on the topological charge $N$ such that the effective mass
$k$ is smaller for the two--vortex and hence its exponential  decay
slower. In Fig. 3 we show the energy density and
electric field of the one and two--vortex.
The coupling here is the lower critical coupling $\kappa_\ix{cr}^\ix{l}$.


\section{Conclusions}


We have studied classical static vortex solutions 
in an $O(3)\, \sigma$ Chern--Simons system with
unbroken $U(1)$ gauge symmetry.
The vortices have an electric charge which shows a 
unique maximum dependent
on the coupling to the gauge field.
The magnetic flux in the model
is effectively quantised in the limit of strong coupling while 
the angular momentum of the vortices is fractional such that they can
be considered as classical anyons.

In the case of two vortices sitting on top of each other,
the model has a repulsive and two attractive phases, 
depending on the parameter which couples the gauge and matter fields.
This has interesting 
consequences for the interaction of multivortices.
In the repulsive regime
they will presumably try to  move away from each other and
for a bounded region this would lead 
to a configuration similar to an Abrikosov--lattice with vortices in
equidistant and fixed positions.
Such configurations occur in the description of 
flux tubes in type--II  superconductors.
In the attractive regime, however, vortices which are not too widely 
separated from each other will be likely to coalesce.
In this context it is worth investigating whether the vortices of higher 
winding number show a similar dependence on the coupling, in particular
whether their critical couplings 
$\kappa_\ix{cr}^\ix{l}$ and 
$\kappa_\ix{cr}^\ix{h}$ 
(if they exist) are of the same value than they are here.

The inter--vortex forces at large and medium distances will be dominated
by the
matter fields, because the electromagnetic fields decay faster by a factor of 
$e^{-kr}$.
Thus, the intercations should be 
well described asymptotically by the dipole
picture developed in \cite{multi}.

\vspace{1cm}
{\Large{\bf Acknowledgement}}\medskip\\
It is a pleasure to thank Wojtek Zakrzewski, Bernard Piette and Jacek Dziarmaga for
stimulating discussions.
I also wish to thank Roman Jackiw
for bringing two references to my attention.
I acknowledge an EPSRC grant No: 94002269.


\newpage

\begin{center}
{\huge {\bf Figure Captions}}
\end{center}
\hspace{0.5cm}
{\large {\bf Fig. 1}}\\
1a) The static energy $E$ (\ref{en}) 
in units of $4\pi N$ as a function of the
Chern--Simons coupling parameter $\kappa$ for $N=1$ (solid line)
and $N=2$ (dotted line).
The plot includes the Bogomol'nyi bound (dashed line).\\
1b) The electric charge $Q$ (\ref{flux_charge_r}) 
in units of $2\pi N$ as a function of the
Chern--Simons coupling parameter $\kappa$ for $N=1$ (solid diamonds) and  $N=2$ (triangles up).\\ 

{\large {\bf Fig. 2}}\\
2a) The energy density $e$ (\ref{led}) as a function of $r$
for $N=1$ and $\kappa = 0.3$ (dotted), 
$\kappa = \kappa_\ix{cr}^\ix{l} = 0.632$ (solid)
and  $\kappa = 2$ (dashed).\\
2b) The profile function $f$ as a function of $r$ for
$N=1$ and $\kappa = \kappa_\ix{cr}^\ix{l} =
0.632$ (solid), $\kappa = 2$ (dashed), 
$\kappa = 50$ (dot--dashed).\\
2c) The gauge field $a$ as a function of $r$ for
$N=1$ and
$\kappa = \kappa_\ix{cr}^\ix{l} = 0.632$ (solid), $\kappa = 2$ (dashed),
$\kappa = 0.4$ (dotted).\\
2d)  The gauge field  $v$ as a function of $r$ for
$N=1$ and $\kappa = \kappa_\ix{cr}^\ix{l}
= 0.632$ (solid),$\kappa = 2$ (dashed), 
$\kappa = 0.4$ (dotted).\\

{\large {\bf Fig. 3}}\\
3a) The energy density $e$ (\ref{led}) as a function of $r$ for
$\kappa = \kappa_\ix{cr}^\ix{l} = 0.632$, $N=1$ (solid) and $N=2$ (dashed).\\
3b) The electric fields radial component $E_r$  as a function of $r$ for
$\kappa  = \kappa_\ix{cr}^\ix{l}= 0.632$, $N=1$ (solid) and $N=2$ (dashed).\\


\begin{thebibliography}{99}



\bibitem{bep}
A. Belavin and A.M. Polyakov: JETP. Lett. 22 (1975)  245
\bibitem{sky}
T.H.R. Skyrme: Proc.Roy.Soc. A260 (1961) 127
\bibitem{multi}
B.M.A.G. Piette, B.J. Schroers and W.J. Zakrzewski: Z.Phys. C65
(1995) 165 \\
B.M.A.G. Piette, B.J. Schroers and W.J. Zakrzewski: Nucl.Phys.
B439 (1995) 205
\bibitem{nar}
G. Nardelli: Phys.Rev.Lett. 73 (1994) 2524
\bibitem{Bernd} B.J. Schroers:  Phys.Lett. B356 (1995) 291
\bibitem{jps} J. Gladikowski, B.M.A.G. Piette and B.J. Schroers:
Phys.Rev. D53 (1996) 844
\bibitem{ghosh} P.K. Ghosh and S.K. Ghosh:
Phys.Lett. B366 (1996) 199
\bibitem{lee1}
K. Kimm, K. Lee and T.Lee: Phys.Rev.  D53 (1996) 4436 
\bibitem{jlw}
S.K. Paul and A. Khare: Phys.Lett. B174  (1986) 420\\
R. Jackiw and E.J. Weinberg: Phys.Rev.Lett. 64 (1990) 2234\\
Hong, Y. Kim and P.Y. Pac: ibid. 2230\\
R. Jackiw, K. Lee and E.J. Weinberg: Phys.Rev. D42
(1990) 3488
\bibitem{llmk}
C.Lee, K.Lee and H.Min: Phys.Lett. B252 (1990) 79.\\
C.Kim: Phys.Rev. D47 (1993) 673
\bibitem{wil}
F. Wilczek: Fractional Statistics and Anyonic Superconductivity.
World Scientific 1990
\bibitem{btw}
M.A. Mehta, J.A. Davis and I.J.R. Aitchison: Phys.Lett. B281 (1992) 86\\
B.M.A.G. Piette, D.H. Tchrakian and W.J. Zakrzewski: Phys.Lett.
B339 (1994) 95\\
D.H. Tchrakian and K. Arthur: Phys.Lett. B352 (1995) 321
\bibitem{hedge}
H. Weigel, B. Schwesinger and G. Holzwarth: Phys.Lett.  B168
 (1986) 321\\
V.B. Kopeliovich and B.E. Stern: Pis'ma v ZhETF45
(1987)  165\\
J.J.M. Verbaarschot, T.S. Walhout, J.Wambach and H.W. Wyld:\\
Nucl.Phys. A468 (1987) 520
\bibitem{der}
R. Hobart: Proc.Phys.Soc.Lond. 82 (1963) 201\\
G.H. Derrick: J.Math.Phys. 5  (1964) 1252




\end{thebibliography}
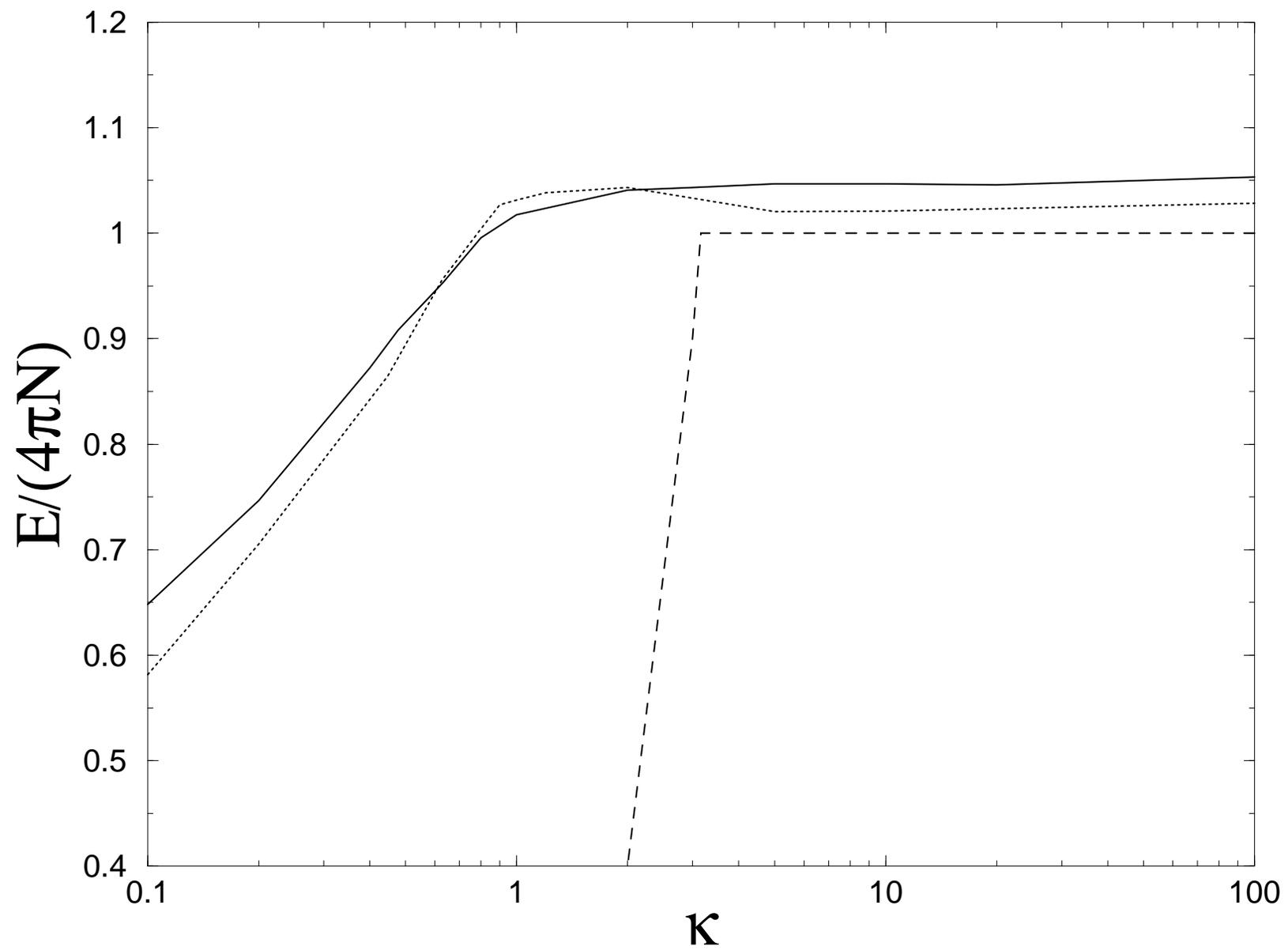
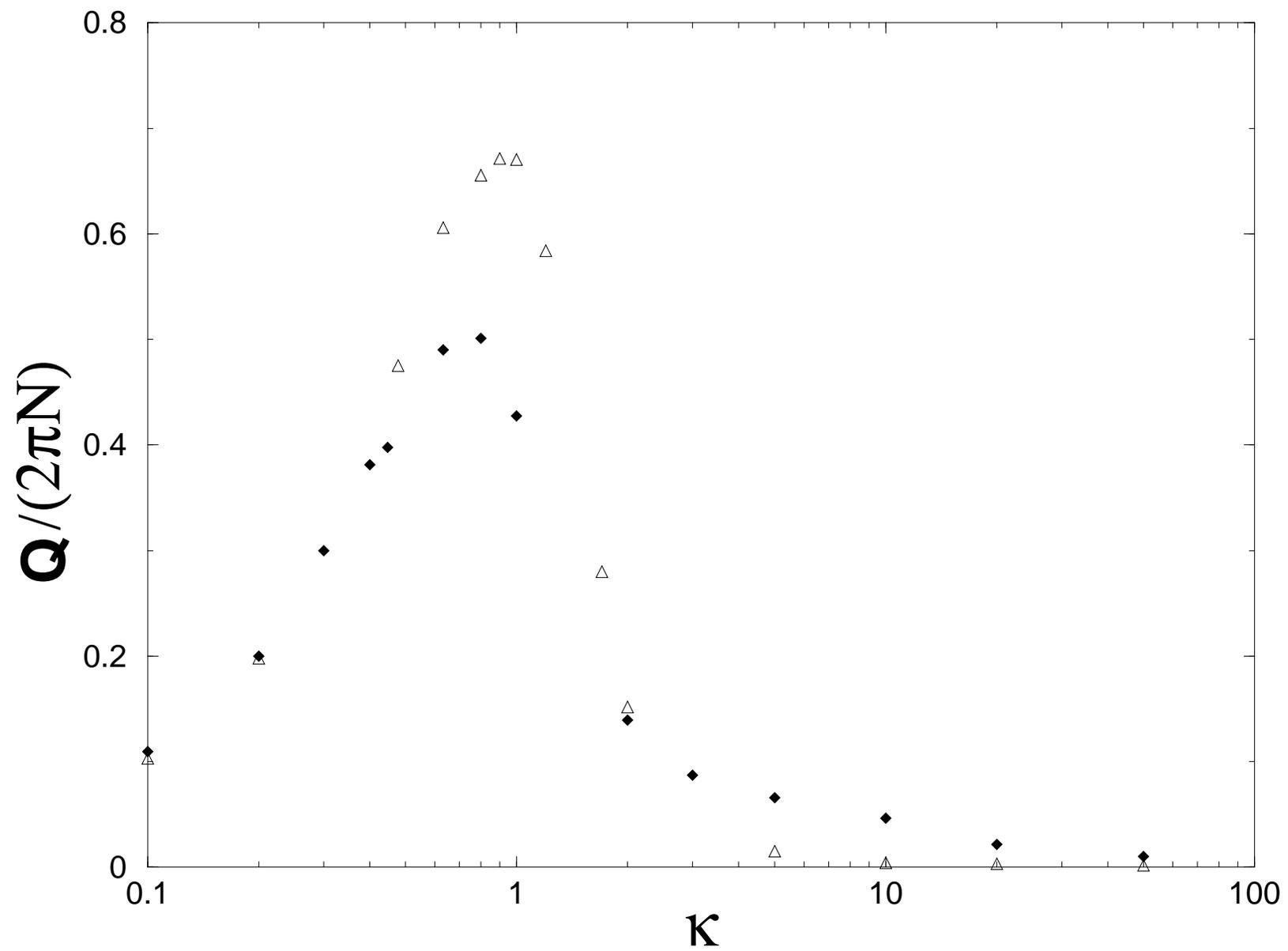
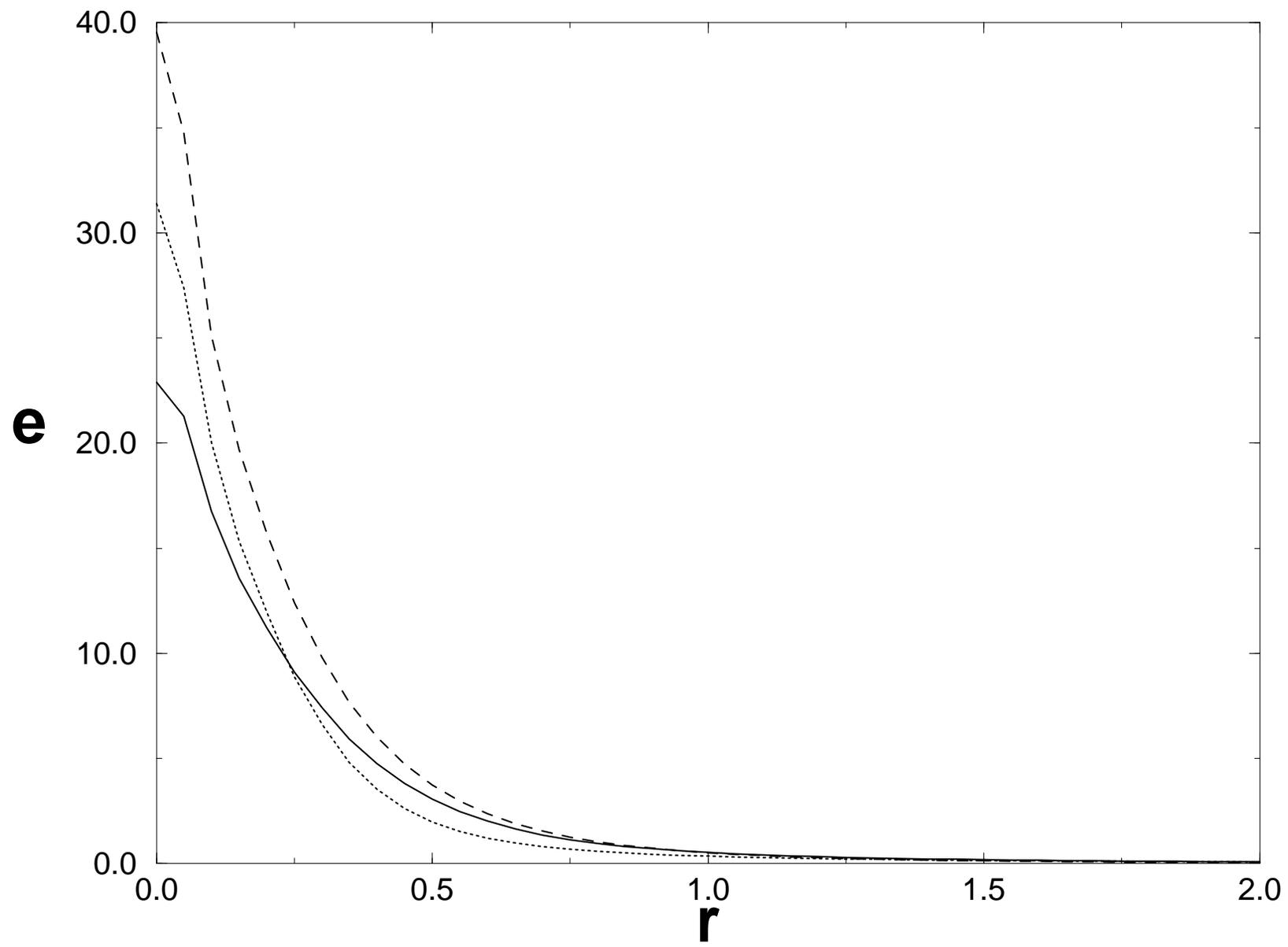
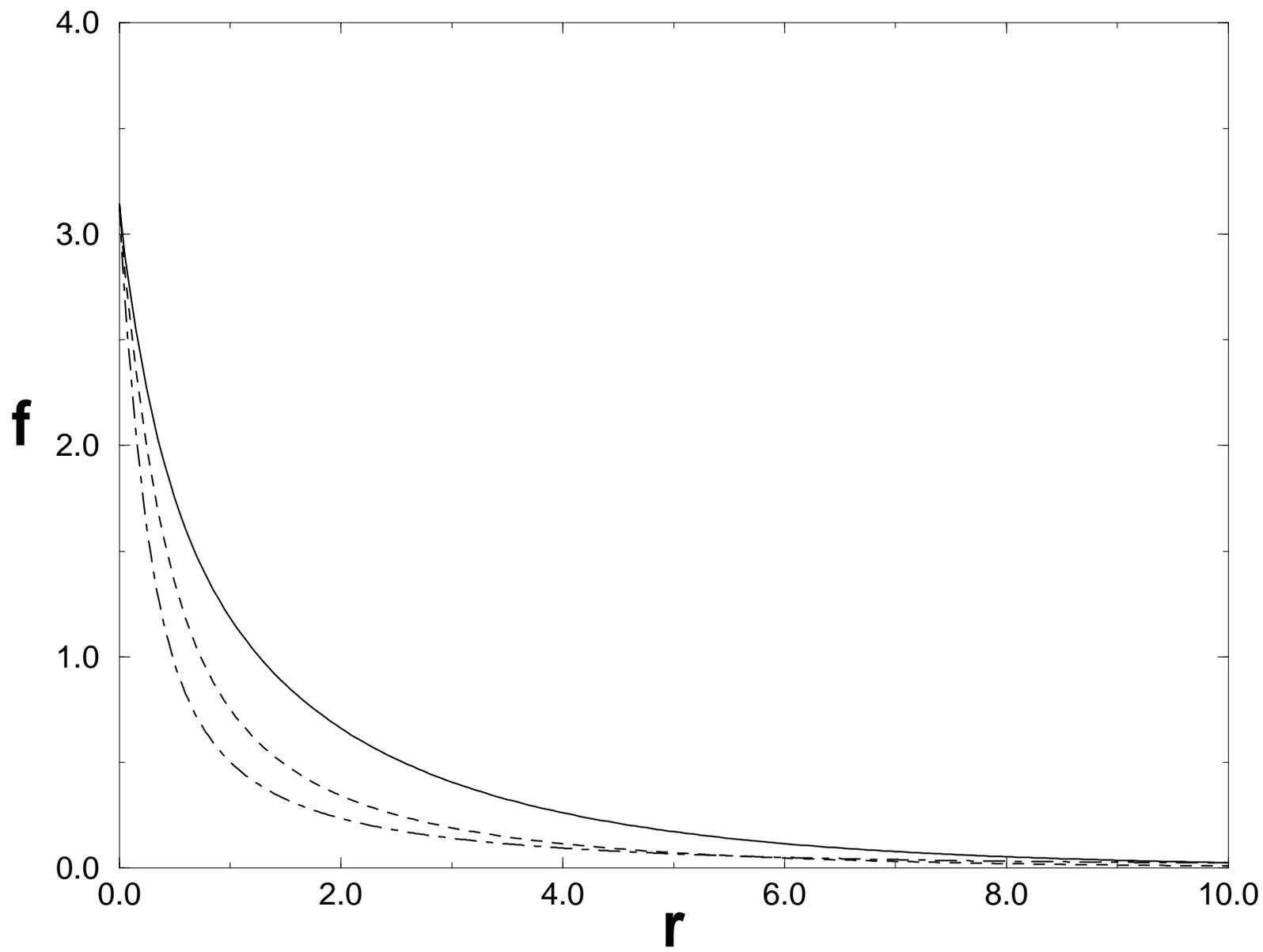
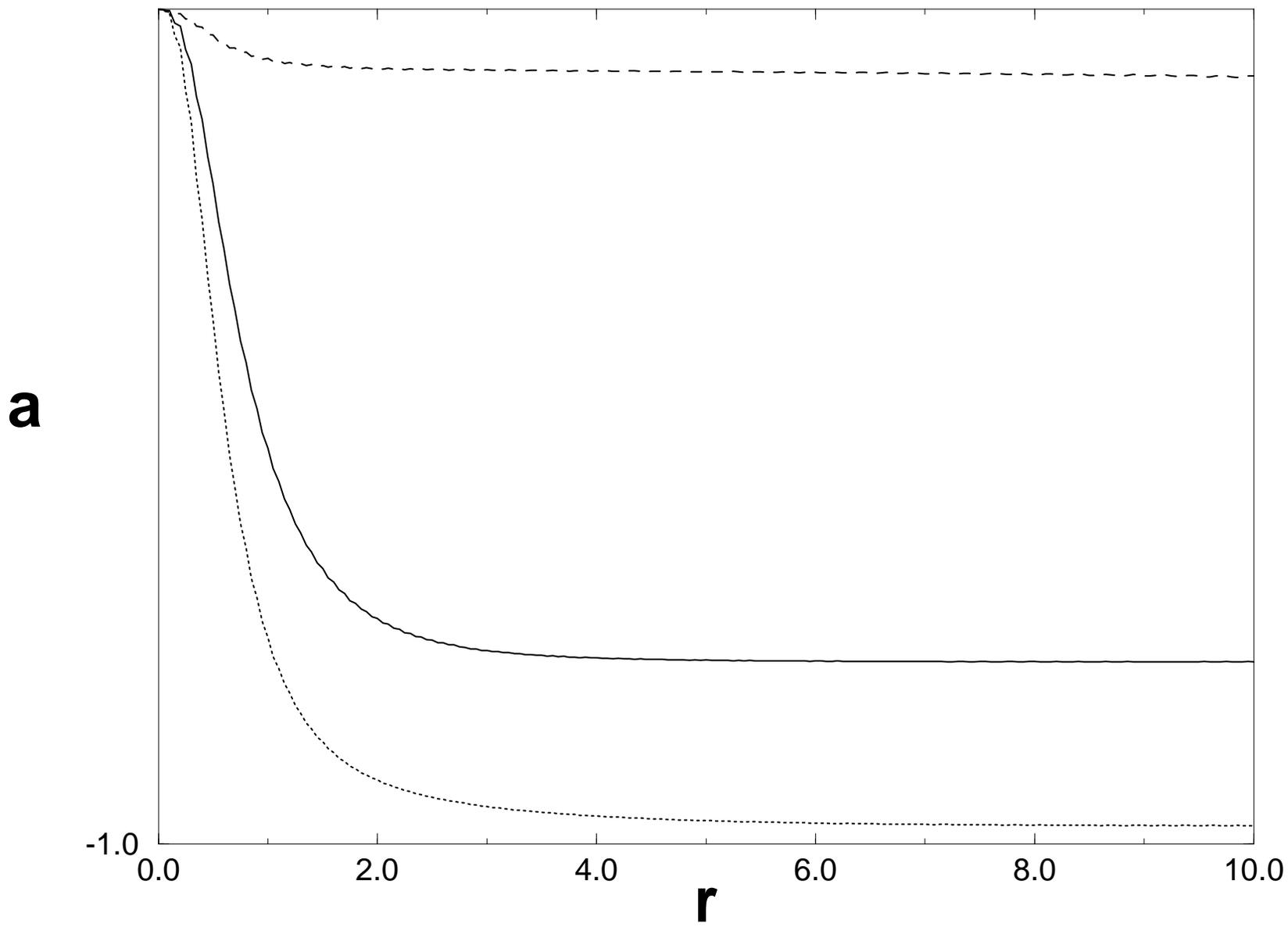
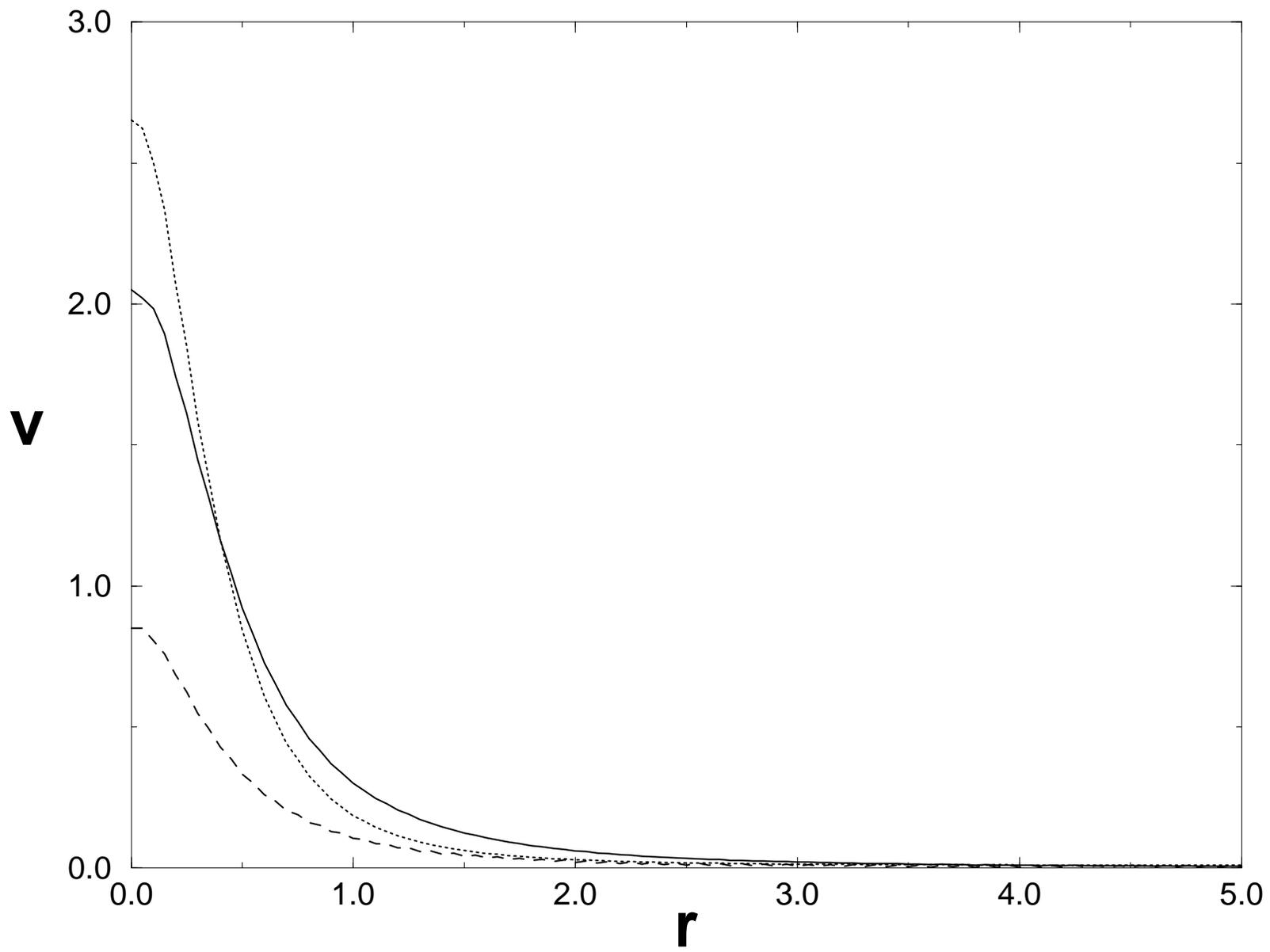
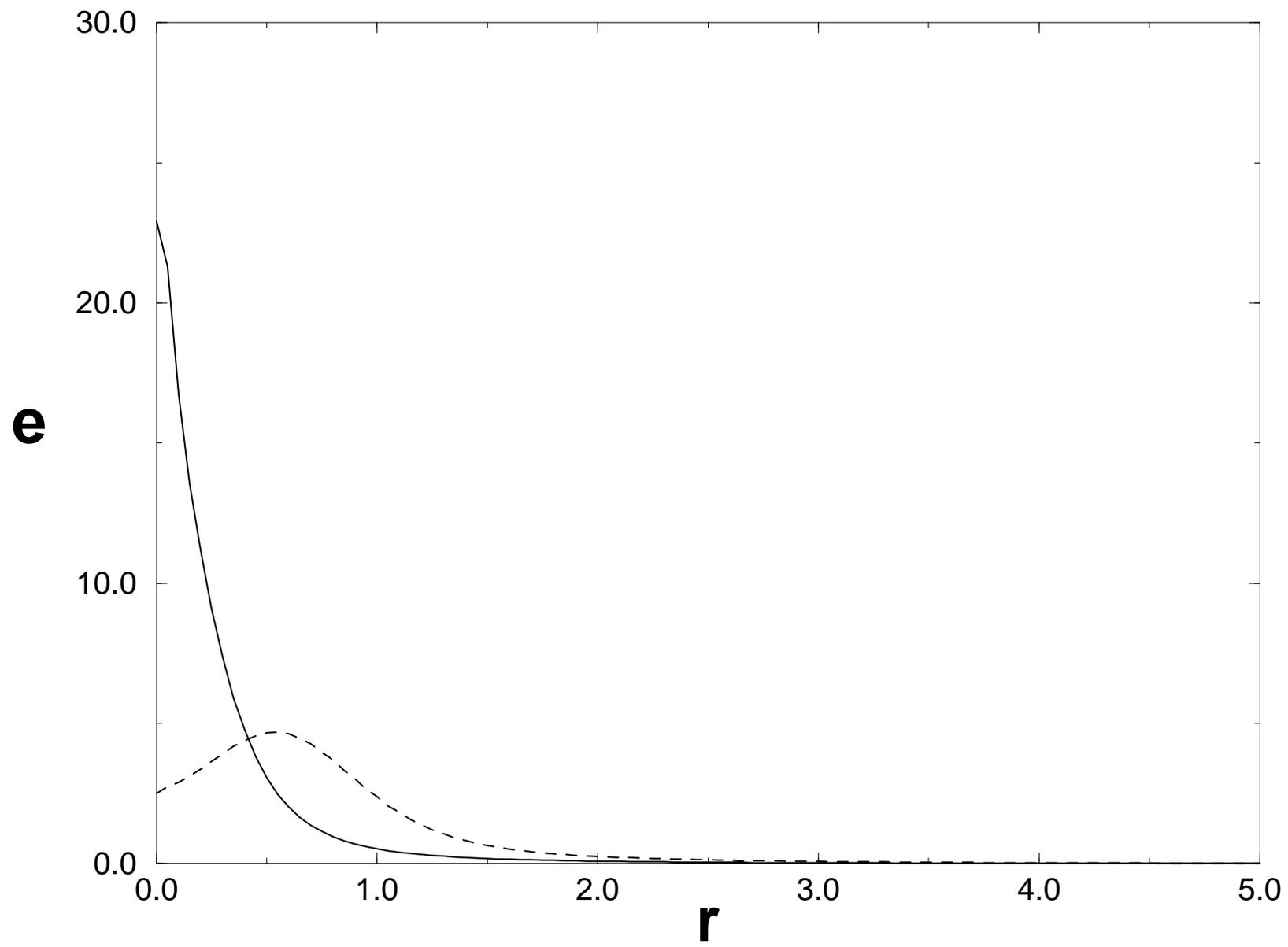
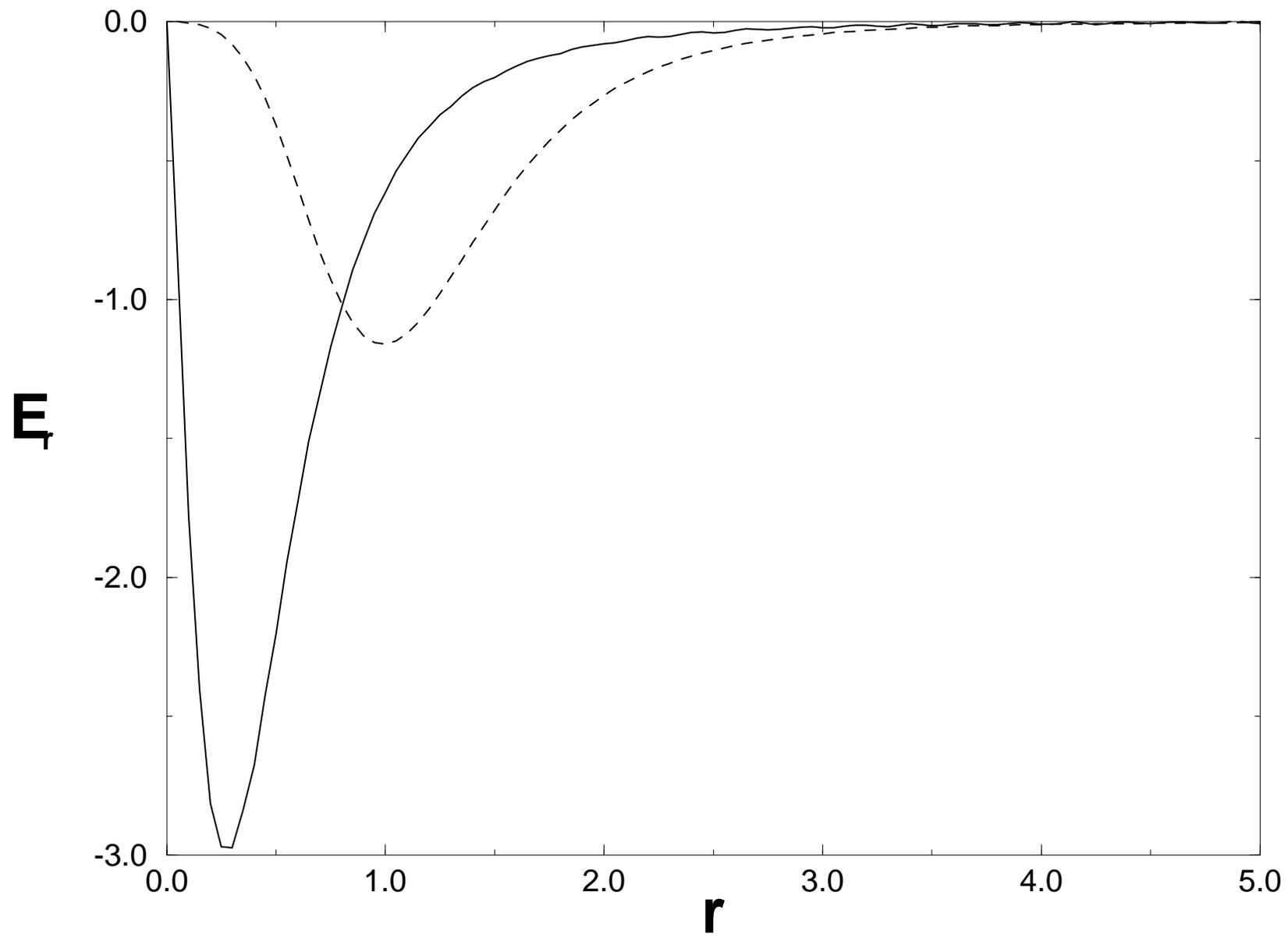
\end{document}